\documentclass[final,3p,times,twocolumn]{elsarticle}

\usepackage{graphicx}

\usepackage{amssymb}

\usepackage{endfloat}

\journal{Computers \& Geosciences}

\begin{document}
\begin{frontmatter}



\title{SEAMONSTER: A Demonstration Sensor Web Operating in Virtual Globes}


\author[UAS]{M. J. Heavner\fnref{MattLANL}}
\ead{matt.heavner@gmail.com}
\author[Vexcel]{D. R. Fatland}
\author[UAS]{E. Hood}
\author[UAS]{C. Connor}

\address[UAS]{University of Alaska Southeast, 11120 Glacier Highway, Juneau, Alaska 99801}
\address[Vexcel]{Vexcel Corporation, Microsoft, Boulder, Colorado}

\fntext[MattLANL]{now at Los Alamos National Lab, MS D436, Los Alamos, NM 87544, heavner@lanl.gov}

\begin{abstract}
  A sensor web is a collection of heterogeneous sensors which
  autonomously reacts to the observed environment.  The SouthEast
  Alaska MOnitoring Network for Science, Technology,
  Education, and Research (SEAMONSTER) project has implemented a
  sensor web in partially glaciated watersheds near Juneau, Alaska, on
  the edge of the Juneau Icefield.  By coupling the SEAMONSTER sensor
  web with digital earth technologies the scientific utility,
  education and public outreach efforts, and sensor web management of
  the project all greatly benefit.  This paper describes the
  scientific motivation for a sensor web, the technology developed to
  implement the sensor web, the software developed to couple the
  sensor web with digital earth technologies, and demonstrates the
  SEAMONSTER sensor web in a digital earth framework.
\end{abstract}


\end{frontmatter}


\section{Introduction}
\label{}

A sensor web is fundamentally a distributed set of sensors coupled
with in-web computational power sufficient to autonomously respond to
observed changes in the environment.  The Juneau area combines the
Juneau Icefield, Tongass National Forest, and Inside Passage in close
proximity.  Diverse sensing needs in the area provide an environment
to demonstrate the sensor web concept and to provide a testbed for
maturing sensor web component technologies.

The long term monitoring of changing characteristics of partially
glaciated watersheds (driven by glacier recession) requires long
duration, low sample rate observations to capture seasonal, annual,
decadal or longer trends.  Conversely, to study the impact of sudden
events in the watersheds (such as catastrophic drainage of
supraglacial lakes or extreme precipitation events) requires much
higher sample rate observations.  The autonomous reconfiguration
between these two sampling regimes illustrates the sensor web concept.
To implement this autonomy, field computers must operate year round in
harsh environments ({\it e.g.} glaciers, the Tongass temperate
rainforest or nunataks bordering the glaciers).  The power constraints
associated with remote computer operation create additional
requirements for sensor web autonomy, namely power management.  The
SEAMONSTER project has implemented a prototype sensor web in partially
glaciated watersheds in Juneau, Alaska on the margin of the Juneau
Icefield.  In order for the diverse data sets gathered by
heterogeneous sensors to be of maximum societal benefit, the data set
must be discoverable and visualizable.  The use of public data archves
or other alternative ways of hosting the data were not used because of
the critical importance of data availability for ``in-web'' use to
trigger autonomous reconfiguration and potential unavailability of
internet connectivity.  The coordinated management of a diverse set of
large numbers of sensors requires considerable infrastructure.  Both
the needs for data exploration and sensor web management can be
combined through the use of virtual globe technologies, as described
in this paper.

\subsection{Sensor Webs}
\label{}

An environmental sensor network is a distributed set of sensors,
generally operating in a mode of storing the data locally and
periodically sending the data via telemetry or manual download.
Communication between nodes, coupled with in-web computational power
provides the sensor network the capability for autonomous
reconfiguration based on the observed environment.  The condition
which triggers the reconfiguration could be of scientific interest or
hazard monitoring and response (such as a glacial lake outburst) or an
operational event (such as a temperature sensor which fails or a
decrease of available battery power below a critical threshold).  The
autonomous reconfiguration of the sensor network is the key feature of
a sensor web.  The Open Geospatial Consortium (OGC) has developed a
set of Sensor Web Enablement (SWE) protocols (OGC reference document
OGC-07-165).  SEAMONSTER is utilizing OGC and OGC SWE protocols and
practices.  The SEAMONSTER project is primarily supported by the NASA
Earth Science Technology Office (ESTO) Advanced Information Systems
Technology (AIST) project.  In the final report from the 2007 NASA
Earth Science workshop, the sensor web concept is defined as ``a
coherent set of heterogeneous, loosely-coupled, distributed nodes,
interconnected by a communications fabric that can collectively behave
as a single dynamically adaptive and reconfigurable observing system.
The Nodes in a sensor web interoperate with common standards and
services.  Sensor webs can be layered or linked together''
\cite{AIST_2007}.  A software description of the sensor web concept is
provided by \cite{Gibbons:2003p1394}.  The sensor web concept
incorporates the need for data discovery for unanticipated data use.
This aspect of sensor webs is for both the incorporation of
unanticipated existing or co-collected data sets as well as the
dissemination of sensor web data sets to interested parties
(incorporation and publication of data).  The SEAMONSTER project has
served as a testbed for both incorporation and publication of data.
Another aspect of sensor webs that is not discussed in this paper is
the integration of model analysis with data sets.  The model and data
integration within the sensor web can quickly help fine tune analyses
and identify anomalous behavior in addition to the reduction and
analysis of data gathered.  The Earth Information Services prototype
demonstrates model integration with the SEAMONSTER sensor web
\cite{Hansen:2009p2273}.

\subsection{SEAMONSTER}
\label{}
SEAMONSTER is the SouthEast Alaska MOnitoring Network for Science,
Technology, Education, and Research project. The SEAMONSTER sensor web
backbone of field-hardened, autonomous power-managing computers has
been developed and deployed in the Lemon Creek watershed in Southeast
Alaska. Lemon Creek is a small, glacial watershed that hosts a
diversity of temperate ecosystems. At the head of the watershed, 1200
m above sea level, lies the Lemon Creek Glacier. The Lemon Creek
Glacier covers approximately 20\% of the watershed with a layer of ice
up to a few hundred meters thick. Following the watershed down from
the mountain peaks surrounding the glacier toward the marine
environment, the watershed encompasses a range of complex and diverse
ecosystems in a fairly small spatial expanse.  The ecosystems include
sparsely vegetated alpine tundra, lush alpine meadows, new and
old-growth temperate rainforests, cold streams, and, at the lowest
reach, tidally-influenced wetlands and an estuary region.  One aspect
of note for this study area is that Lemon Creek Glacier is part of the
Juneau Icefield Research Program (JIRP) and was monitored during
International Geophysical Year (IGY) (1957-58) \cite{Miller:1999p1290}
and continuously on through the International Polar Year (IPY)
(2007-8), providing a relatively long-term record of watershed
changes.  Some of the information gathered by JIRP is published as kml
datasets and can be viewed in parallel with the kml publications from
SEAMONSTER, illustrating the power of virtual globes to easily
integrate publicly available datasets from disparate sources via kml
standards.

The integration of the SEAMONSTER sensor web with virtual globes was
originally inspired by the James Reserve network sensor network
integration with Google Earth \cite{Askay_2006}.  The SEAMONSTER
virtual globe effort was focused on outreach (as was the James Reserve
network).  The SEAMONSTER efforts in virtual globes presented in this
paper expand on the James Reserve work by including the integration of
the geowiki with the postgres database and the geoserver.  The
additional integration of tools enhances the virtual globes'
contributions to collaboration, development, and technology
dissemination efforts.

\subsection{Science Motivation}
\label{}

\begin{figure}[p]
  \begin{picture}(480,200)
    \put(0,0){\resizebox{400pt}{!}{\includegraphics*{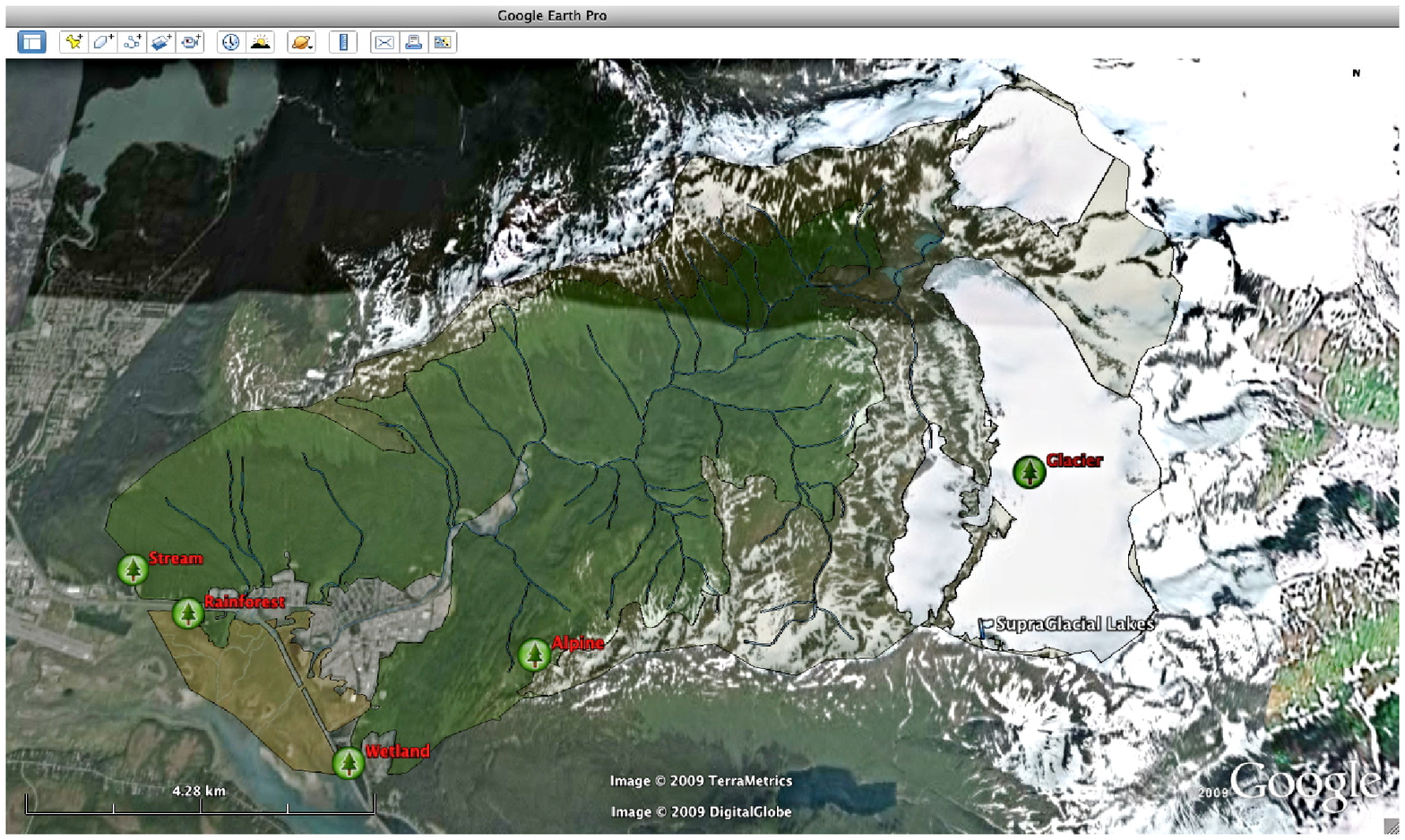}}}
    \end{picture} \caption{The Lemon Creek Watershed.  The figure illustrates .shp files exported from ESRI's Arc* suite of software, served by the PostGIS powered GeoServer, used to designate the different ecosystem portions of the Lemon Creek Watershed.  The supraglacial lake forms on the south end of the Lemon Creek Glacier (which flows towards the North, which is up in this view).}
    \label{fig:SM_site}
\end{figure}

Glaciers in southeastern Alaska have been retreating and thinning
rapidly for the last several decades
\cite{Larsen:2007p472,Arendt:2002p1}.  This loss of ice and the
associated increase in freshwater discharge has important implications
for the hydrology of pro-glacial rivers and the physical properties in
downstream receiving marine ecosystems \cite{Hood:2008p869}. The
proximity to the Juneau Icefield, the fifth largest icefield in North
America, allows for the relatively easy deployment of a multi-layered
sensor web to address fundamental questions regarding the ice dynamics
and hydrology of outlet glaciers draining the icefield.  Lemon Creek
Glacier ($\sim$10~km$^2$) has a single supraglacial lake which fills
during the summer and catastrophically drains into Lemon Creek, a
relatively well constrained glacial hydrologic system, illustrated in
(Figure~\ref{fig:SM_site}).

SEAMONSTER has begun expanding into the Mendenhall Glacier watershed.
Approximately a dozen ice-marginal lakes form on the Mendenhall
Glacier. Mendenhall Glacier terminates in Mendenhall Lake. These two
features make the glacial hydrology of Mendenhall more complicated
than the Lemon Creek system. The critical question the deployed sensor
web will address is: Are accelerated rates of glacial melt and the
associated increased runoff beneath the glacier creating a positive
feedback effect by increasing the rate of basal ice motion which, in
turn, increases the rate at which ice is being delivered to low
elevation ablation zones?  What in turn, is the impact of changing
glacial control of the watersheds on the ecosystem?  Understanding the
relationship between glacial hydrology and glacier mass balance is
critical for predicting environmental responses to climate change.

\section{Methods}
\label{}

The primary technology goal of the SEAMONSTER project is to
instantiate a testbed sensor web in a harsh environment with multiple
relevant science use cases.  Hardware efforts were driven by
requirements for power, communications, and in-situ processing required
for autonomy.  Software requirements included autonomous
reconfiguration, data management, data discoverability, data browsing,
and integration of additional, unanticipated data streams.  Design
goals for both hardware and software included the integration of
existing technology with clear documentation of any modifications made
for the SEAMONSTER project.

\subsection{Hardware}

The backbone of the SEAMONSTER sensor web is a small headless field
computer or ‘Microserver’ developed by Vexcel Corporation in Boulder
Colorado. Vexcel Microservers were developed from 2003 to 2008 with
NASA support. These were conceived as general-purpose sensor platforms
spanning signal frequencies from one sample per day up to kilohertz
sampling frequency, with applicability across a broad variety of field
science disciplines and applications (seismology, meteorology, visual
monitoring, GPS surveys of glacier motion, robotic surveys of stream
and lake chemistry and more). Microservers are field-hardened for
survivability in harsh environments. They include a standard COTS WiFi
router or an equivalent high bandwidth communication device to enable
creation of ad hoc field networks used for sharing data and in the
future enabling sophisticated software to direct limited field
resources towards interesting events. Microservers are also able to
act as base stations for localized lightweight sensor networks built
on mote technology such as TelosB motes available from Crossbow
Technologies. Microservers address the power-cost-data challenge in
environmental monitoring by providing a battery recharging system and
a power conditioning subsystem that permits the unit to hibernate when
the primary external energy source (typically a lead-acid battery) is
depleted. Hibernation continues (drawing microwatts of power) until
the system determines that the primary battery is sufficiently
recharged, typically by solar panels. The computing environment is an
ARM-based Single Board Computer and the device also incorporates a GPS
board. Supported communication protocols in addition to WiFi include a
PC/104 bus, Ethernet, USB, and RS-232 serial ports as well as several
analog-to-digital (ADC) channels. Data storage capacity of many
Gigabytes is provided by a solid state USB flash drive.

As noted Microservers can act as base stations for second-tier
lightweight mote-based sensor networks. SEAMONSTER follows the
technology lead of the Johns Hopkins University “Life Under Your Feet”
soil ecology program employing the Koala / FCP protocols to deploy
‘physical heartbeat’ sensors: Total Solar Radiation,
Photosynthetically Active Radiation, temperature, soil moisture,
relative humidity, and electrical conductivity. The motes use the
802.15.4 Zigby protocol to periodically recover data to the base
station (Microserver file system) and from there a series of daemons
move raw data to the SEAMONSTER online GeoServer data catalog.

\subsection{Sensors}

Weather stations, based on Campbell Scientific data loggers, have been
deployed at the top of Lemon Creek Glacier (near the lake) and near
the terminus of the glacier to measure parameters such as air
temperature, precipitation, solar radiation, wind speed and direction,
snow depth, {\it etc}. The station overlooking the lake includes a
high-resolution digital pan/tilt/zoom web camera.  A pressure
transducer to measures lake level is installed in the supra-glacial
lake. Mounted downstream in Lemon Creek is an YSI probe that measures
water-quality characteristics such as water temperature, pH, dissolved
oxygen, and turbidity.

\subsection{Software}

\begin{figure}[p]
  \begin{picture}(480,240)
    \put(0,0){\resizebox{400pt}{!}{
        \includegraphics*{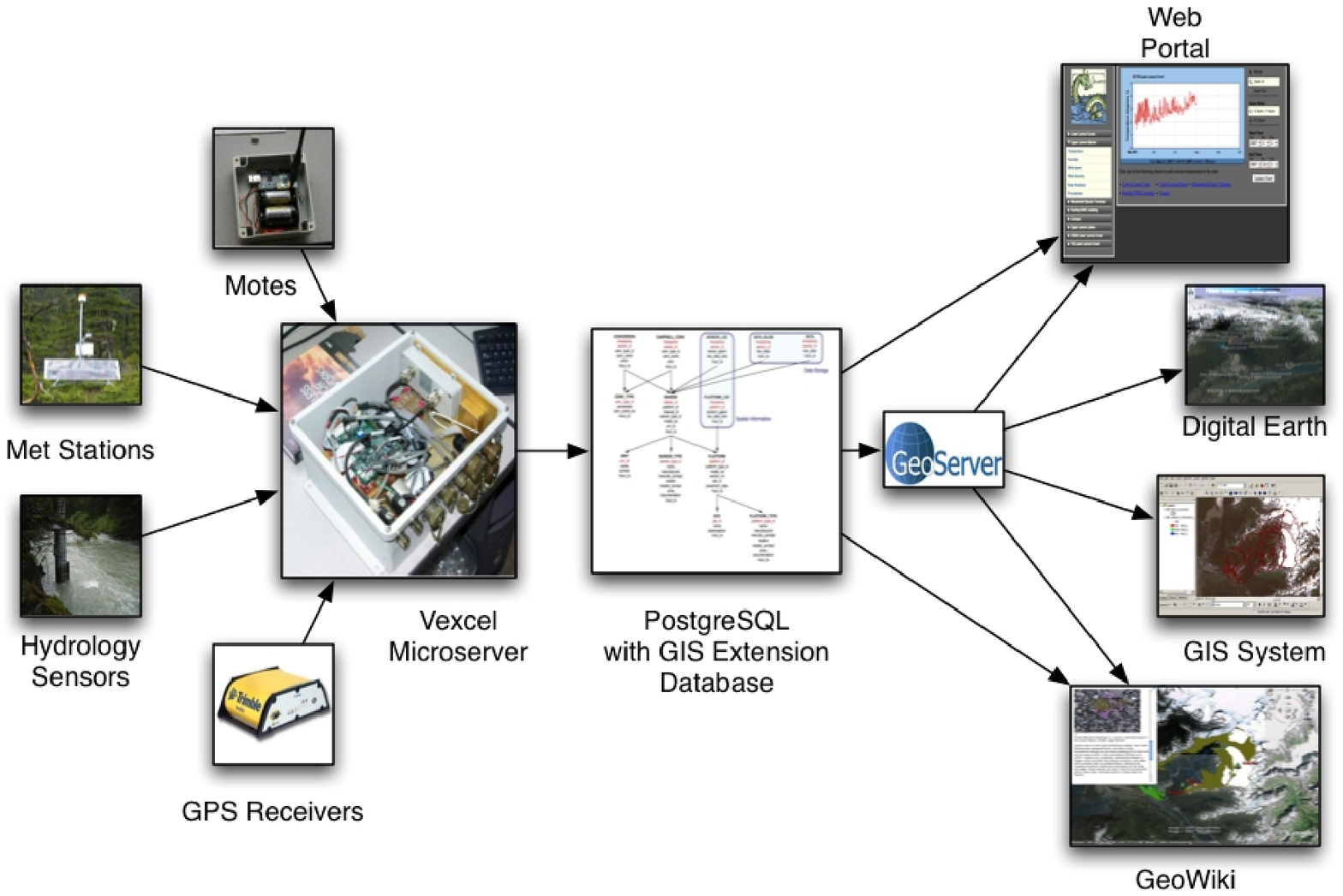}}}
    \end{picture} \caption{Conceptual Diagram showing sensor flow to Vexcel Microserver, aggregation in the PostGIS database, and multiple output streams.}
    \label{fig:SM_DataFlow}
\end{figure}

The SEAMONSTER project generates heterogeneous data sets at irregular
time intervals.  Managing the data with ease of access, public
outreach, and easy of comparison between the different instruments for
researchers motivated our use of a single SQL database for final
storage of all data.  (In situ data is typically stored as ASCII files
within the Microserver filesystem.)  As illustrated in
Fig.~\ref{fig:SM_DataFlow}, all the data streams through sensors to
the microservers and into a postgreSQL database with GIS extensions
enabled, called PostGIS \cite{Hsu:2007p1594}.  The GIS extensions
require that every table entry be associated with a location and time
entry.  Coupling the PostGIS database with the Open Geospatial
Consortium (OGC) GeoServer automatically provides the ability to
disseminate the data streams through a web portal
(http://seamonster.jun.alaska.edu/browser/), kml for 4-D Geobrowsers
(such as Google Earth or Microsoft Virtual Earth), services to more
traditional GIS systems (such as the ESRI suite of Arc* software), and
through the geowiki.  The PostGIS database stores both raster and
vector ({\it e.g.} ESRI .shp files).  The geowiki is a project wiki
using mediawiki, which requires a SQL database for page storage.  By
using the same PostGIS database as SEAMONSTER data, a spatio-temporal
location is required for each wiki page.  Fig.~\ref{fig:SM_DataFlow}
illustrates the SEAMONSTER architecture with output to end-users.  For
readability, the data feedback loops enabling the sensor web aspects
of SEAMONSTER are not shown, but the feedback loops are an integral
aspect of the SEAMONSTER sensor web.

To enable autonomous reactivity, SEAMONSTER has implemented several
strategies.  The Vexcel Microservers continuously run on-board code
used to react to the local power conditions.  Vexcel Microservers have
scheduled scripts running to analyze data from other platforms and
react.  For example, when the pressure transducer data indicates the
supraglacial lakes begin to drain, the scripts on the Vexcel
Microserver force the pan/tilt/zoom camera to image the draining lake.
SEAMONSTER also serves as a testbed platform for more sophisticated
sensor web management solutions such as the MACRO (Multi-agent
Architecture for Coordinated, Responsive Observations) project
implementing a CORBA-based solution
\cite{JohnSKinnebrew:2007p1458}. For resource management, SEAMONSTER
makes use of the munin/rrdtool package and integrates the scheduled
collection and plotting of resource information in the kml generation.

\section{Results}
\label{}

The coupling of the PostGIS database and GeoServer allows multiple
dynamically generated access methods, as illustrated on the right side
of Fig~\ref{fig:SM_DataFlow}.  This section describes three different
examples of the access methods to the SEAMONSTER sensor web with
virtual globes.  The first example describes the geowiki, primarily
illustrating the education and public outreach benefits of coupling
sensor webs and virtual globes through Google Earth network links and
dynamically generated kml.  The data browsing and access potential of
coupling SEAMONSTER and virtual globes is illustrated in the second
subsection describing the integration of the SEAMONSTER data browser
and Microsoft's Virtual Earth via openlayers technology.  Finally, the
benefits of using virtual globe technology for sensor web management
are described.

\subsection{Geowiki}

\begin{figure}[p]
  \begin{picture}(480,240)
    \put(0,0){\resizebox{400pt}{!}{
        \includegraphics*{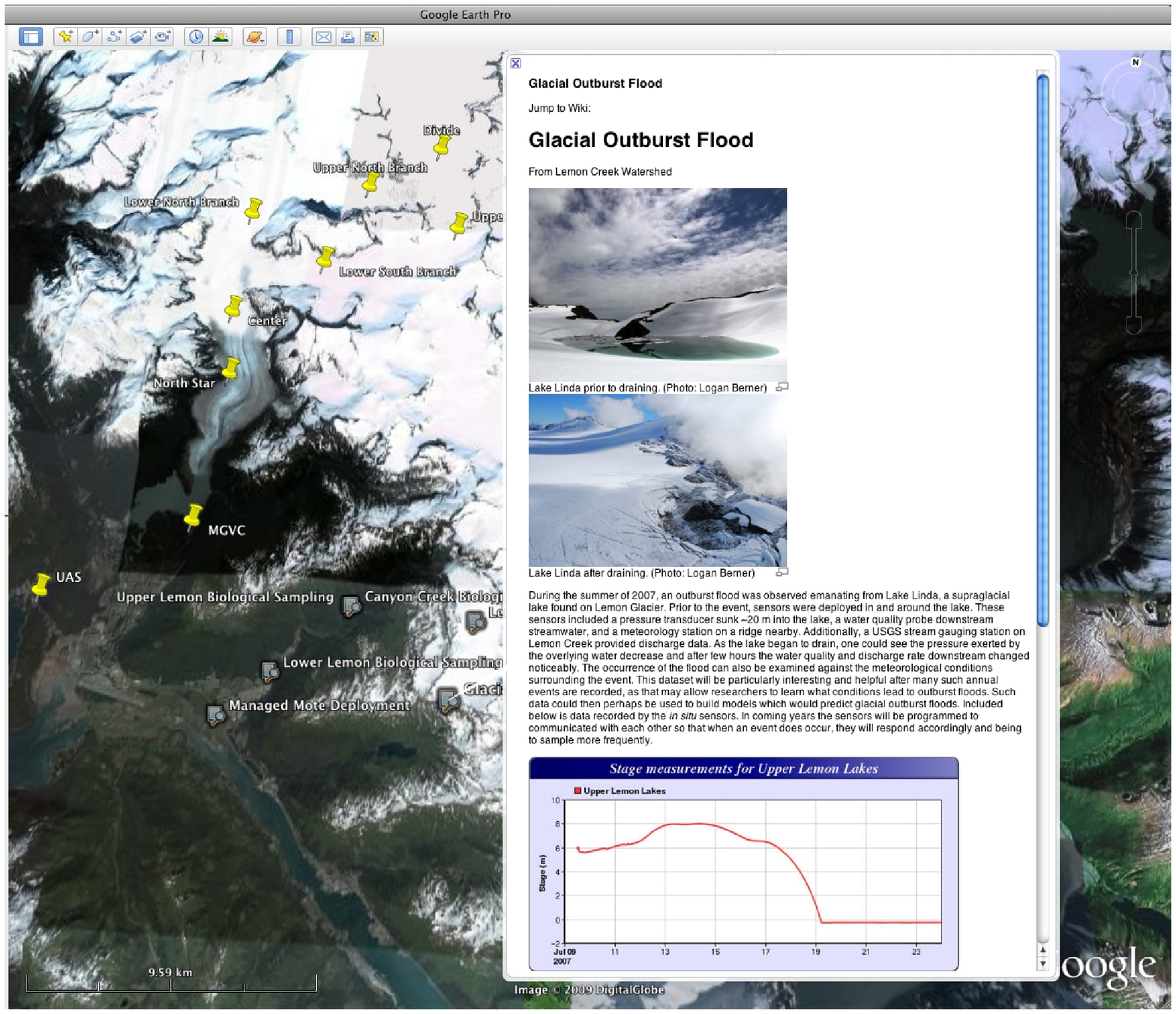}}}
  \end{picture} \caption{A screencapture of the dynamically generated SEAMONSTER geowiki content, primarily used for education and public outreach.  The Mendenhall Glacier and Lemon Creek measurements sites are shown as push pins, geowiki content pages describing various features of the Lemon Creek Watershed are shown, with the Glacial Outburst Flood page opened, showing data observed during the 2007 lake drainage.}
    \label{fig:Geowiki_GE}
\end{figure}

The SEAMONSTER sensor web is testing new technology development for
NASA, documenting impacts of climate change and glacier control of
watersheds, and providing views and information about a popular
tourist destination (approximately one million tourists visit Juneau
every year).  Education and public outreach are a major component of
the SEAMONSTER project \cite{Berner:2007p1711}.  As part of the
efforts for education and public outreach as well as facilitating data
discovery and sharing by other scientists, a public wiki was conceived
of as a two-way conduit for general information about the sensor web.
The wiki is intended to have a larger scope: a hypothetical example is
a biologist interested in wind and temperature data in the Juneau area
for a migratory bird study.  The SEAMONSTER meteorologic data could be
discovered and the migratory bird information could be easily added by
the biologist to the public wiki.  The mediawiki engine is used to
implement the public wiki, requiring a SQL database backend.
Typically, mediawiki is configured to use mySQL.  However, the
SEAMONSTER mediawiki installation makes use of the PostGIS database
already containing the SEAMONSTER data.  The benefit of using the
PostGIS database in conjunction with the public wiki is the PostGIS
requirement that every database entry have geospatial information--so
every wiki page is georeferenced.  Fig.~\ref{fig:Geowiki_GE}
illustrates the geowiki, with a screen capture from Google Earth
showing the output from a network link, dynamically generated by the
GeoServer requests into the PostGIS database. The network link
provides both the kml features and the geowiki page content.  The
geowiki network links provide the wiki pages within Google Earth (the
Glacial Outburst Flood wiki page is shown, along with the dataset
showing the lake drainage at the bottom of the wiki page view).
Another geowiki network link provides ecosystem polygons generated by
ArcView (as .shp files) which delineate glacier, wetland, stream,
alpine, and rainforest areas of the Lemon Creek watershed.  Over fifty
photographs related to the watershed are included as a layer served by
the GeoServer/PostGIS geowiki.  The integration of the virtual globe,
sensor web, and wiki technologies through the PostGIS database,
GeoServer, network links, and kml provides the SEAMONSTER education
and public outreach portal.

\subsection{Data Access}

\begin{figure}[p]
  \begin{picture}(480,240)
    \put(0,0){\resizebox{400pt}{!}{
        \includegraphics*{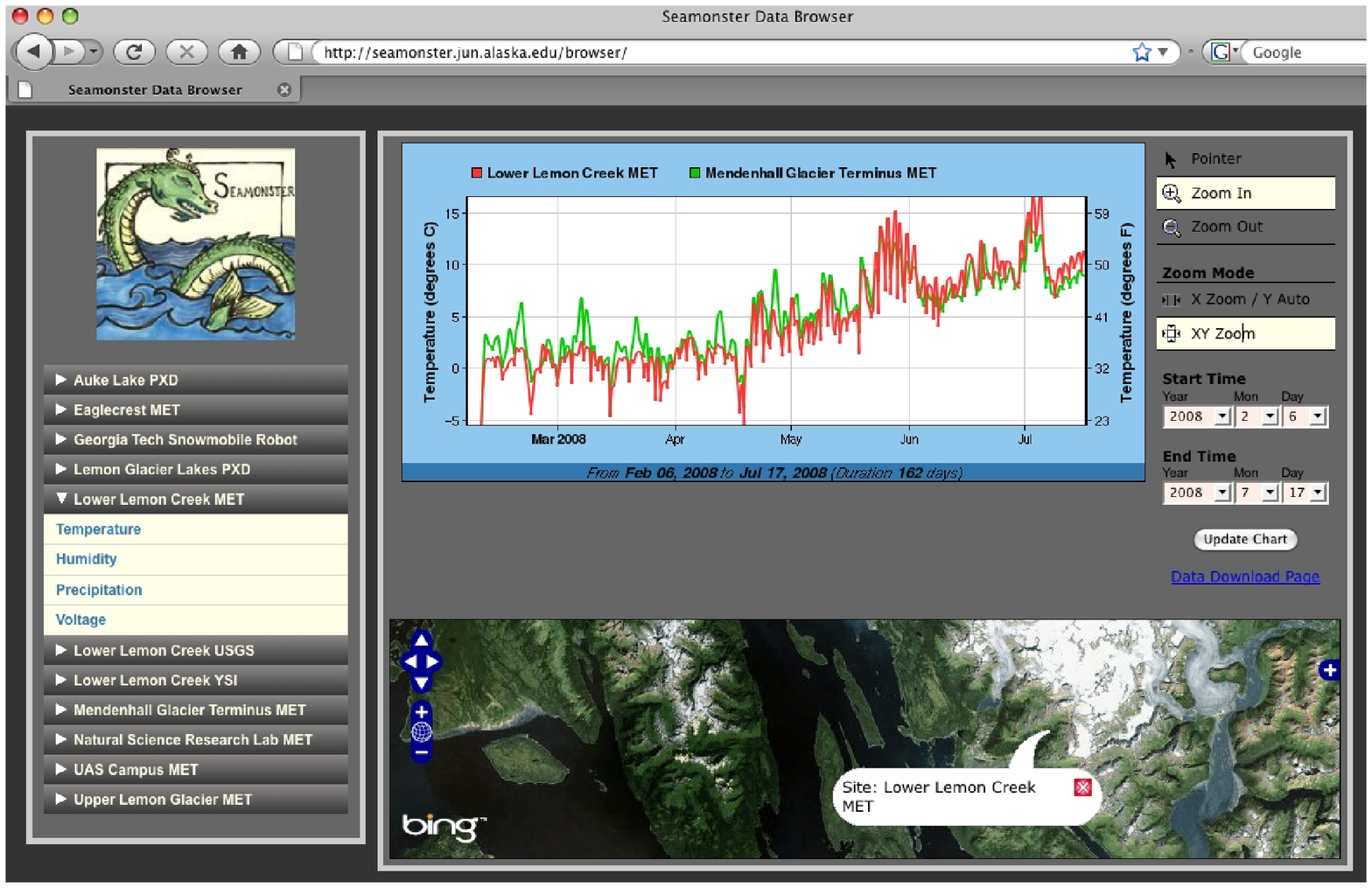}}}
    \end{picture} \caption{A screenshot from the dynamically generated SEAMONSTER Data Browser (http://seamonster.jun.alaska.edu/browser/).  This provides data browsing capability, including the ability to compare similar measurements from diverse locations, in the example shown the temperature at the Mendenhall Glacier terminus (green) is compared with the temperature at the Lower Lemon Creek stations (red).  The station location is shown in the Microsoft Bing geobrowser using openlayers.  The SEAMONSTER data portal is dynamically generated from the same PostGIS database as seen in the previous geowiki example.}
    \label{fig:DataBrowser}
\end{figure}

The diverse data sets and non continuous sampling from various sites
created data cataloging and browsing issues.  A web portal is used to
avoid creating ''one-off'' in-house solutions which may not be
available for interested public or scientists.
Fig.~\ref{fig:DataBrowser} illustrates two temperature records from
different sites in the SEAMONSTER study area.  The left panel shows a
list of all the stations in gray.  Selecting a station triggers (via
openlayers) the map at the lower portion of Fig.~\ref{fig:DataBrowser}
to identify the location of the sensor.  The measurements available at
the station are shown as a drop down menu (Temperature, Humidity,
Precipitation, and Voltage are shown).  After selection of a
measurement, the graph updates to show the most recent year of
measurements.  The user can zoom in and out of the plot graphically or
select a date range.  One more dataset of a similar measurement from a
different location can be overplotted on the plot, as is shown in
Figure~\ref{fig:DataBrowser}.  The raw data can be downloaded in
various formats (text, netcdf, xls) via this web interface.  The
SEAMONSTER data browser shown in Fig.~\ref{fig:DataBrowser} is a
second example of the benefits of the integration of the Sensor Web
and Virtual Globe concepts.  The data browser code is available
through the project SVN (see Section~\ref{section:resources}) and has
been tested for use by at least two other projects.

\subsection{Sensor Web Management}

\begin{figure}
  \begin{picture}(480,400)
    \put(0,0){\resizebox{400pt}{!}{
        \includegraphics*{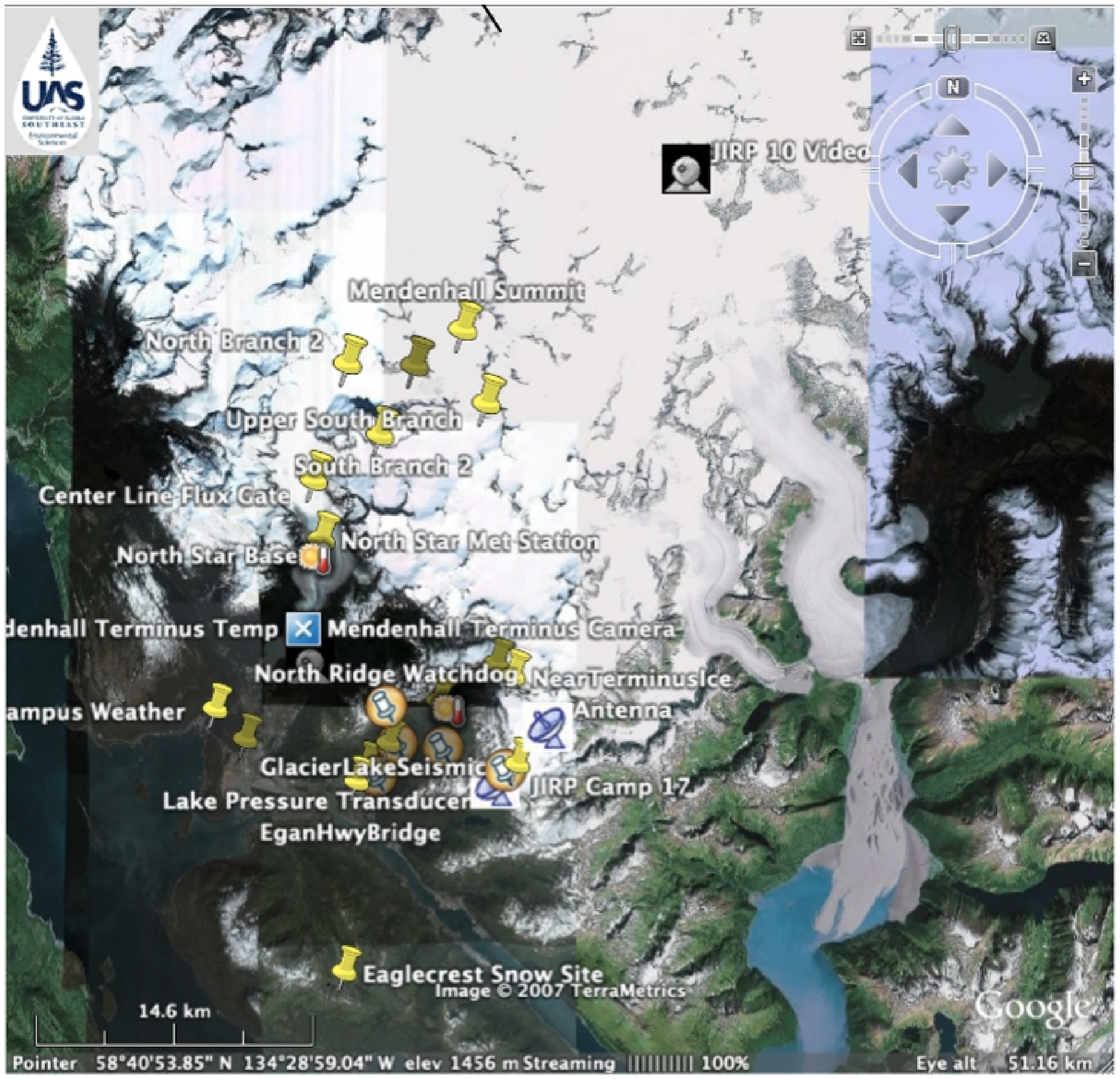}}}
    \end{picture} \caption{A screenshot from the dynamically generated SEAMONSTER kml file.  This file is generated based on the status of various sensor web platforms.  For example, the data from Mendenhall Glacier Terminus site has not been recently updated, so it is flagged with a large blue X icon, clearly visible among the diverse icons representing sensors, cameras, antennae, {\it etc}.}
    \label{fig:SM_Management}
\end{figure}

The third benefit resulting in the integration of the sensor web and
virtual globe technologies is in the operation and management of the
diverse resources of the sensor web.  Visualizing power,
communication, and other state of health information is critical,
especially in the distributed, relatively harsh environment in which
SEAMONSTER is operating.  Fig.~\ref{fig:SM_Management} illustrates a
large number of placemarks designating sensors (Lake Pressure
Transducer), locations (Eaglecrest Snow Site), future instrument
placement (JIRP 10 Video), and other components of the sensor network
(Antenna, North Ridge Watchdog).  Even in this relatively busy view,
it is reasonably easy to identify the Mendenhall Terminus Camera as
the large blue X placemarker.  This is used to indicate that data from
the camera has not been received in the time period it was expected.
Fig.~\ref{fig:SM_Management} illustrates the SEAMONSTER sensor web
management kml file which is regularly generated and embeds system
information collected via munin
and is useful in diagnosing system problems (for example, munin
collects and plots battery voltage information, which is attached to
the placemarks illustrated in the plot).

\section{Future Work}

This paper describes the technology developed for integration of the
SEAMONSTER sensor web and virtual globes.  The SEAMONSTER project has
developed and documented hardware and software.  Rigorous evaluation
of the success of the virtual globe technology and end user testing
has not been undertaken.  Feedback from technologists, scientists, and
collaborators has been positive, but has only been collected and acted
upon at the anecdotal level.  A full evaluation of the SEAMONSTER
virtual globe interface for usability is beyond the scope of the
project.  However, the technology developed for SEAMONSTER integration
with virtual globes has been utilized as a prototype by other data
dissemination groups and has been enhanced and subjected to user
testing.

\section{Conclusions}
\label{}

The integration of sensor web and virtual globe technology provides
dramatic benefits to the goals of sensor web (more efficient
observations) and virtual globe (visualization and understanding)
visions.  The three examples provided illustrate the education/public
outreach, data discovery and exploration, and sensor web management
benefits realized by the coupling of sensor web and virtual globe
technologies.  The hardware and software used by SEAMONSTER to
instantiate a sensor web coupled with virtual globes is described in
hopes that the benefits realized by the SEAMONSTER effort may be
duplicated (and improved upon).  A final section provides pointers to
the specific resources described.

\section{Resources}
\label{section:resources}

The integration between sensor web and digital earth has been done in
an open environment, coordinated through a project wiki, with all code
stored in a publicly accessible subversion code repository.

The SEAMONSTER public {\bf geowiki} is available at

\begin{center}http://seamonsterak.com/\end{center}

and is intended to be primarily useful for
the description of the SEAMONSTER study area.  The {\bf SEAMONSTER
  project wiki} at 

\begin{center}http://robfatland.net/seamonster\end{center}

is used to document the development efforts and technologies used with
sufficient detail to replicate the sensor web and virtual globe
efforts.  All of the code developed by SEAMONSTER (from assembly code used
for power control boards to SQL database design and php code for the
data browser) is available through the {\bf SEAMONSTER SVN} at

\begin{center}http://seamonster.jun.alaska.edu/websvn/.\end{center}

  The {\bf SEAMONSTER
  databrowser} is available at

\begin{center}http://seamonster.jun.alaska.edu/browser/.\end{center}

\section{Acknowledgements}
\label{}

Funding for SEAMONSTER is provided through NASA Earth Science
Technology Office grant AIST-05-0105, NOAA Education Partnership Panel
Interdisciplinary Scientific Environmental Technology (ISET)
Cooperative Science Center Grant, and NSF Research Experience for
Undergraduates Grant No. 0553000. Marijke Habermann, Logan Berner,
Edwin Knuth, Nick Korzen, David Sauer, Josh Galbraith, Shannon
Siefert, and Nathan Rogers have been integral to the SEAMONSTER
project.


\begin{thebibliography}{10}

\bibitem[{AIS(2007)}]{AIST_2007}
NASA AIST Workshop, Feb 2007. {NASA} {AIST} {S}ensor {W}eb {T}echnology {M}eeting {R}eport.
  Http://esto.nasa.gov/sensorwebmeeting/files/
  AIST\_Sensor\_Web\_Meeting\_Report\_2007.pdf.

\bibitem[{Arendt et~al.(2002)Arendt, Echelmeyer, Harrison, Lingle, and
  Valentine}]{Arendt:2002p1}
Arendt, A., Echelmeyer, K., Harrison, W., Lingle, C., Valentine, V., 2002.
  Rapid wastage of Alaska glaciers and their contribution to rising sea level.
  Science 297~(5580), 382--386.

\bibitem[{Askay(2006)}]{Askay_2006}
Askay, S.~P., Dec 2006, New Visualization Tools for Environmental Sensor Networks: Using Google Earth as an Interface to Micro-Climate and Multimedia Datasets. Univ. California Riverside Masters Thesis.

\bibitem[{Berner et~al.(2007)Berner, Habermann, Hood, Fatland, Heavner, and
  Knuth}]{Berner:2007p1711}
Berner, L., Habermann, M., Hood, E., Fatland, R., Heavner, M., Knuth, E., Jan
  2007. Providing a virtual tour of a glacial watershed. EOS Trans, AGU
  88~(52).

\bibitem[{Gibbons et~al.(2003)Gibbons, Karp, Ke, Nath, Seshan, Res, and
  Pittsburgh}]{Gibbons:2003p1394}
Gibbons, P., Karp, B., Ke, Y., Nath, S., Seshan, S., Res, I., Pittsburgh, P.,
  2003. Irisnet: An architecture for a worldwide sensor web. IEEE Pervasive
  Computing 2~(4), 22--33.

\bibitem[{Hansen et~al.(2009)Hansen, LeFebvre, Schultz, Romberg, Mysore, Holub,
  McCaslin, Salm, Esterline, Li, Baber, Fuller, Pogue, Wright, Heavner,
  Steinbach, Olobode, Qian, and Fatland}]{Hansen:2009p2273}
Hansen, T.~L., LeFebvre, T.~J., Schultz, M., Romberg, M., Mysore, A., Holub,
  K., McCaslin, P., Salm, S., Esterline, A., Li, Y., Baber, C., Fuller, K.,
  Pogue, Y., Wright, W., Heavner, M., Steinbach, M., Olobode, R., Qian, L.,
  Fatland, R., 2009. Earth information services. 25th Conference on
  International Interactive Information and Processing Systems (IIPS) for
  Meteorology, Oceanography, and Hydrology Session 7B, Internet Applications
  and Cyberinfrastructure II.

\bibitem[{Hood and Scott(2008)}]{Hood:2008p869}
Hood, E., Scott, D., Sep 2008. Riverine organic matter and nutrients in
  southeast Alaska affected by glacial coverage. Nature Geoscience 1~(9), 583--587.

\bibitem[{Hsu and Obe(2007)}]{Hsu:2007p1594}
Hsu, L., Obe, R., 2007. PostGIS for geospatial analysis and mapping. Postgres
  OnLine Journal, 19--20.

\bibitem[{Kinnebrew et~al.(2007)Kinnebrew, Biswas, Shankaran, Schmidt, and
  Suri}]{JohnSKinnebrew:2007p1458}
Kinnebrew, J.~S., Biswas, G., Shankaran, N., Schmidt, D.~C., Suri, D., April
  2007. Integrating task allocation, planning, scheduling, and adaptive
  resource management to support autonomy in a global sensor web. 2007 {NASA}
  Science Technology Conference.

\bibitem[{Larsen et~al.(2007)Larsen, Motyka, Arendt, and
  Echelmeyer}]{Larsen:2007p472}
Larsen, C., Motyka, R., Arendt, A., Echelmeyer, K., Jan 2007. Glacier changes
  in southeast Alaska and northwest British Columbia and contribution to sea
  level rise. Journal of Geophysical Research 112~(F01007).

\bibitem[{Miller and Pelto(1999)}]{Miller:1999p1290}
Miller, M., Pelto, M., Dec 1999. Mass balance measurements on the Lemon Creek
  Glacier, Juneau Icefield, Alaska, 1953-1998. Geografiska Annaler 81,
  671--681.

\end{thebibliography}

%
\end{document}